\documentclass[%
 reprint,
superscriptaddress,
 amsmath,amssymb,
 aps,
]{revtex4-2}

\usepackage{amssymb,amstext,amsmath}
\usepackage{graphicx}
\usepackage{dcolumn}
\usepackage{bm} 
\usepackage{bbm} 
\usepackage{multirow}
\usepackage{bigstrut}
\usepackage{makecell,rotating}
\usepackage{mathrsfs}
\usepackage{booktabs}
\usepackage{threeparttable}
\usepackage{subfigure}
\usepackage{chngpage}
\usepackage{float}
\usepackage{color}
\usepackage{mathtools}
\usepackage{url}
\usepackage{hyperref}
\usepackage{booktabs}
\usepackage{tabularx}
\hypersetup{
    colorlinks=true,
    linkcolor=blue,
    citecolor=green,
    urlcolor=black,
    linkbordercolor=0 0 1
}

\definecolor{darkred}{rgb}{0.75,0,0}
\definecolor{darkgreen}{rgb}{0,0.5,0}
\definecolor{darkblue}{rgb}{0,0,0.75}
\definecolor{darkorange}{rgb}{0.8,0.3,0}
\definecolor{dark}{rgb}{0,0,0}

\begin{document}

\preprint{APS/123-QED}

\title{Evolution of cooperation and competition in multilayer networks}

\author{Wenqiang Zhu}
\thanks{These authors contributed equally to this work.}
\affiliation{%
School of Artificial Intelligence, Beihang University, Beijing 100191, China}
\affiliation{%
Key Laboratory of Mathematics, Informatics and Behavioral Semantics, Beihang University, Beijing 100191, China}
\affiliation{%
Zhongguancun Laboratory, Beijing 100094, China}

\author{Xin Wang}
\thanks{These authors contributed equally to this work.}
\affiliation{%
School of Artificial Intelligence, Beihang University, Beijing 100191, China}
\affiliation{%
Key Laboratory of Mathematics, Informatics and Behavioral Semantics, Beihang University, Beijing 100191, China}
\affiliation{%
Zhongguancun Laboratory, Beijing 100094, China}
\affiliation{%
Hangzhou International Innovation Institute, Beihang University, Hangzhou 311115, China}
\affiliation{%
Beijing Advanced Innovation Center for Future Blockchain and Privacy Computing, Beihang University, Beijing 100191, China}
\affiliation{%
State Key Laboratory of Complex \& Critical Software Environment, Beihang University, Beijing 100191, China}

\author{Chaoqian Wang}%
\affiliation{%
Department of Computational and Data Sciences, George Mason University, Fairfax, VA 22030, USA}

\author{Weijie Xing}
\affiliation{%
School of Artificial Intelligence, Beihang University, Beijing 100191, China}
\affiliation{%
Key Laboratory of Mathematics, Informatics and Behavioral Semantics, Beihang University, Beijing 100191, China}

\author{Longzhao Liu}
\affiliation{%
School of Artificial Intelligence, Beihang University, Beijing 100191, China}
\affiliation{%
Key Laboratory of Mathematics, Informatics and Behavioral Semantics, Beihang University, Beijing 100191, China}
\affiliation{%
Zhongguancun Laboratory, Beijing 100094, China}
\affiliation{%
Beijing Advanced Innovation Center for Future Blockchain and Privacy Computing, Beihang University, Beijing 100191, China}
\affiliation{%
State Key Laboratory of Complex \& Critical Software Environment, Beihang University, Beijing 100191, China}

\author{Hongwei Zheng}
\affiliation{%
Beijing Academy of Blockchain and Edge Computing, Beijing 100085, China}

\author{Jingwu Zhao}
\thanks{Corresponding author: zhaojingwuruc@163.com}
\affiliation{%
School of Law, Beihang University, Beijing 100191, China}

\author{Shaoting Tang}
\thanks{Corresponding author: tangshaoting@buaa.edu.cn}
\affiliation{%
School of Artificial Intelligence, Beihang University, Beijing 100191, China}
\affiliation{%
Key Laboratory of Mathematics, Informatics and Behavioral Semantics, Beihang University, Beijing 100191, China}
\affiliation{%
Zhongguancun Laboratory, Beijing 100094, China}
\affiliation{%
Hangzhou International Innovation Institute, Beihang University, Hangzhou 311115, China}
\affiliation{%
Beijing Advanced Innovation Center for Future Blockchain and Privacy Computing, Beihang University, Beijing 100191, China}
\affiliation{%
State Key Laboratory of Complex \& Critical Software Environment, Beihang University, Beijing 100191, China}
\affiliation{%
Institute of Medical Artificial Intelligence, Binzhou Medical University, Yantai 264003, China}


\begin{abstract}
Cooperation and competition coexist and coevolve in natural and social systems. Cooperation generates resources, which in turn, drive non-cooperative competition to secure individual shares. How this complex interplay between cooperation and competition shapes the evolution of social dilemmas and welfare remains unknown. In this study, we introduce a two-layer evolutionary game model, in which one layer is a cooperative public goods game, and the other is a competitive involution game, with cross-layer feedback linking the two. We find that feedback can either promote or inhibit cooperation, depending on the baseline conditions. For example, moderate resource and synergy factor values can promote social welfare when feedback strength is large. This provides an approach to adjusting the strength and asymmetry of cross-layer feedback to promote cooperation and social welfare. We thus emphasize the importance of managing feedback mechanisms to balance cooperation and competition in complex social systems.

\end{abstract}

\maketitle

\section{Introduction}
Interactions among individuals in social, biological, and ecological systems involve a balance between cooperation and competition. Cooperation enables individuals to share resources and engage in mutually beneficial behaviors~\cite{trivers1971evolution}, while competition drives individuals to outperform others in securing limited resources~\cite{smith1973logic}. These dynamics are present across various levels of organization, from individual interactions to social structures. Understanding how they coexist and influence one another is crucial to explaining the evolution of social systems.

Evolutionary game theory~\cite{weibull1997evolutionary,nowak2006evolutionary} provides a framework to analyze the emergence and persistence of cooperation on networks~\cite{nowak1992evolutionary,lieberman2005evolutionary,ohtsuki2006simple,perc2013evolutionary,allen2017evolutionary,wang2024evolutionary}. In the public goods game (PGG)~\cite{szabo2002phase,semmann2003volunteering,santos2008social,zhu2024evolutionary,wang2024spatial}, cooperators contribute to a common pool at a personal cost, while defectors benefit from the equal returns without contributing, pushing the system toward the ``tragedy of the commons''~\cite{hardin1968tragedy}. Various mechanisms, including spatial reciprocity~\cite{nowak1992evolutionary,nowak2006five,su2019spatial}, direct and indirect reciprocity~\cite{nowak1998evolution,schmid2021unified,romano2022direct,xia2023reputation,zhu2024reputation}, reward and punishment~\cite{sigmund2001reward,perc2017statistical,zhu2023effects}, environmental feedbacks~\cite{wang2020eco, jiang2023nonlinear}, have been explored to address this issue. On the other hand, competitive interactions, modeled by the involution game (IG)~\cite{wang2021replicator}, reveal how meaningless competition in environments with fixed resources can lead to suboptimal outcomes. Here, the payoff to each participant depends on the utility of one's individual effort in all their efforts~\cite{wang2022modeling,wang2022involution,huang2024memory,chaocheng2023involution,li2024involution,zuo2025simulating}. Rational participants will try to put in more effort to get a larger share, but the overall benefits are not increased. This ``over-competition'' dynamics can be found in economic or organizational settings where employees compete intensely for limited funding~\cite{lazear2000performance}.

Despite extensive research into cooperation and competition, a critical gap remains regarding how cooperation and competition interact and influence one another. Recent studies on multichannel games~\cite{donahue2020evolving,basak2024evolution} have demonstrated that connections among parallel two-person games can facilitate cooperation; however, these studies focus solely on cooperative interactions. Moreover, the mixed cooperative-competitive game model~\cite{lyu2024evolution} investigates intra-group cooperation and inter-group competition but lacks an exploration of feedback mechanisms between them. In real-world systems, cooperation in one context can increase resources, which intensifies competition in another context. For instance, in microbial biofilms, bacteria cooperatively produce extracellular polymeric substances to protect the community while competing for spatial positioning and nutrient resources~\cite{nadell2016spatial}. Similarly, within technology companies, team members collaborate on product, benefiting collectively from improved performance, while competing fiercely due to hierarchical structures and limited opportunities for advancement~\cite{birkinshaw2005intrafirm}. This interplay can drive innovation but may also result in negative effects such as diminished collective output or reduced morale if competitive pressures become excessively high. 

Multilayer networks~\cite{wang2015evolutionary,su2022evolution,de2023more} provide a framework for capturing distinct interactions in separate layers while allowing their interdependence. For example, two-layer networks can separate interaction and strategy update processes to study the effects of structural asymmetry~\cite{ohtsuki2007breaking,chen2021adaptive,inaba2023evolution,wang2023conflict,wang2023greediness}. Other common research introduces additional layers to reflect different scenarios, including referee dynamics~\cite{shi2022two,ling2025supervised}, relationship evolution~\cite{xiong2024coevolution,yue2025coevolution}, intervention strategies~\cite{guo2023third,yan2024inter}, and opinion dynamics~\cite{amato2017interplay}. Furthermore, various features of multilayer networks, such as density heterogeneity~\cite{wang2023public}, link overlap~\cite{battiston2017determinants}, fitness coupling~\cite{wu2021evolution,wang2024utility} and migration~\cite{nag2020cooperation} have been demonstrated to influence the evolution of cooperation.

Building upon these insights, this study proposes a unified two-layer evolutionary game framework of cooperative (PGG) and competitive (IG) dynamics. Cooperative successes increase resources that fuels competition, while intense competition inhibits cooperation. Through Monte Carlo simulations, we explore the conditions under which cooperation and competition promote or inhibit. Our results show that moderate resource availability and synergy factor values promote cooperation across layers, whereas extreme values induce defection. We observe critical transitions between high- and low-cooperation when feedback is strong. Moreover, adjusting the strength and asymmetry of cross-layer feedback can balance cooperation and competition. Optimal coupling enhances cooperation and social welfare, while mismatched feedback may weaken it. These findings provide insight into managing coupled competitive-cooperative systems.

The remainder of this paper is organized as follows. Section~\ref{sec2} introduces the two-layer game model with bidirectional feedback. Section~\ref{sec3} gives the simulation results and analyzes the impact of cross-layer feedback on social dilemmas and social welfare. Finally, section~\ref{sec4} summarizes the conclusions and put forwards the direction of future work.

\begin{figure}[t]
    \centering
    \includegraphics[width=0.5\textwidth]{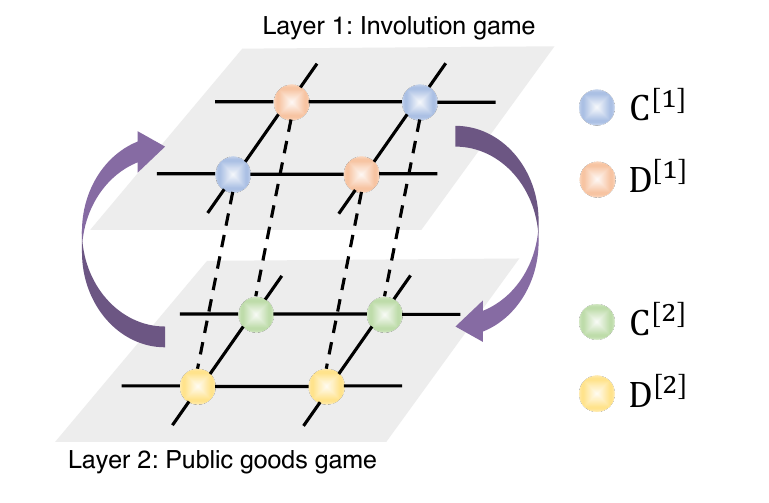}
	\caption{\textbf{Model schematic}. Players are arranged on two interconnected square lattices. Each node represents an individual, with solid lines indicating interactions within layers and dashed lines connecting individuals across layers. The first layer is a competitive IG, and the second layer is a cooperative PGG. Players adopt strategies (C or D) on each layer ([1] or [2]) independently. Purple arrows illustrate that interactions on one layer influence the environment of the other.} 
	\label{model}
\end{figure}

\section{Model}\label{sec2}
We consider a population on two coupled network layers (Fig.~\ref{model}). Each layer is an $L \times L$ periodic square lattice, and each individual occupies a node in both layers. The two layers have identical node positions. An individual $i$ thus engages in two concurrent games: a competitive game on layer 1 and a cooperative game on layer 2. We refer to layer 1 as the IG layer (a competitive environment) and layer 2 as the PGG layer (a cooperative environment). Players select either cooperation (C) or defection (D) strategies for each network layer independently. The strategy adopted by player $i$ at time $t$ on layer $l$ ($l\in\{1,2\}$) is denoted as $S_i^{[l]}(t)$, where $S_i^{[l]}(t) = 1$ indicates cooperation and $S_i^{[l]}(t) = 0$ indicates defection.

Every player engages in games with their nearest $N$ neighbors ($N=5$, assuming von Neumann neighborhood) and calculates payoffs for both IG and PGG layers. We denote by $\mathcal{N}_{\mathrm{C},i}^{[l]}(t)$ ($\mathcal{N}_{\mathrm{D},i}^{[l]}(t)$) the number of players adopting strategy C (D) at time $t$ in the group centered on player $i$ in layer $l$.

\subsection{Payoff calculation in the IG}
In the IG layer, players compete for a baseline resource $M$ by investing different effort levels. Participants choose either a lower-effort strategy (cooperate) with cost $c_1$, or a higher-effort strategy (defect) with cost $d$ ($d > c_1$). The resources that an individual obtains depends on their effort relative to the group's total effort. Following Ref.~\cite{wang2021replicator}, we introduce a relative utility parameter $\beta$ to measure effort effectiveness. Specifically, $\beta = 1$ denotes equal utility across strategies, $\beta > 1$ indicates higher effort yields greater utility, and $\beta < 1$ indicates lower utility.

In the traditional case of no cross-layer effects, the resource $M$ is fixed for each group. In this work, we allow cross-layer feedback: the amount of effective resource in an IG group increases if the corresponding players on layer 2 cooperate more, and vice versa. Let the effective resource for group $i$ on layer 1 be $M^*_i(t)$, 
\begin{equation}\label{feedback_1}
    M_{i}^{*}(t)=M\left[1+\delta\left(\frac{\lambda \mathcal{N}_{\mathrm{C}, i}^{[2]}(t)}{N}-1\right)\right],
\end{equation}
where $0\leq \delta \leq 1$ characterizes the feedback strength from PGG to IG, and $\lambda >1$ represents the amplification effect of local environment. At $\delta=0$, we have $M_{i}^{*}(t)=M$, and the IG layer reduces to the traditional involution game model~\cite{wang2021replicator}. At $\delta=1$, we have $M_{i}^{*}(t)=\lambda \mathcal{N}_{\mathrm{C}, i}^{[2]}(t) /N \cdot M$, and the effective resource completely depends on the feedback from the PGG layer. Moreover, at $\lambda=2$, we have a normalized situation that ensures $M(1-\delta)\leq M_{i}^{*}(t)\leq M(1+\delta)$ when $0\leq \mathcal{N}_{\mathrm{C}, i}^{[2]}(t)\leq N$.

Based on the effective resource $M_{i}^{*}(t)$, the payoffs for a player (C or D, layer 1) in group $i$ at time $t$ are calculated as~\cite{wang2021replicator}
\begin{subequations}
    \begin{align}
        \pi_{\mathrm{C},i}^{[1]}(t)&=\frac{c_1}{\mathcal{N}_{\mathrm{C},i}^{[1]}(t)c_1+\mathcal{N}_{\mathrm{D},i}^{[1]}(t)\beta d}M_{i}^{*}(t)-c_1,\\
        \pi_{\mathrm{D},i}^{[1]}(t)&=\frac{\beta d}{\mathcal{N}_{\mathrm{C},i}^{[1]}(t)c_1+\mathcal{N}_{\mathrm{D},i}^{[1]}(t)\beta d}M_{i}^{*}(t)-d.
    \end{align}
\end{subequations}

Each player accumulates payoffs from participation in $N$ IGs: one centered on themselves and the rest on other neighbors. The total payoffs of player $i$ in layer 1 are
\begin{equation}\label{Pi_lay1}
    \Pi_i^{[1]}(t)=
        \begin{cases}
            \sum_{j\in\Omega_i^{[1]}}\pi_{\mathrm{C},j}^{[1]}(t),&\quad \mathrm{if}~ S_i^{[1]}(t)=1,\\[6pt]
            \sum_{j\in\Omega_i^{[1]}}\pi_{\mathrm{D},j}^{[1]}(t),&\quad \mathrm{if}~ S_i^{[1]}(t)=0,
        \end{cases}
\end{equation}
where $\Omega_i^{[l]}$ denote the set of neighbors of player $i$ in layer $l$, including player $i$ itself.

\begin{table}[t]
    \caption{Main parameters used in this work.}
    \label{tab:symbols}
    \centering
    \renewcommand{\arraystretch}{1.1}    
    \begin{ruledtabular}
    \begin{tabular}{lc} 
        \textbf{Symbol} & \textbf{Interpretation} \\ 
        \midrule
        $L$                 & The side length of the square.                                           \\ 
        $N$                 & The group size.                                          \\ 
        $k$            & The noise in strategy updates.                                                  \\ 
        $c_1$                 & The cost of cooperation in IG .                              \\ 
        $c_2$                 & The cost of cooperation in PGG.                                              \\ 
        $d$                 & The cost of defection in IG.                                      \\
        $M$                 & The baseline of social resources in IG.                          \\ 
        $r$             & The baseline of synergy factors in PGG.                                     \\ 

        $\beta$             & The relative utility of defection in IG.                                        \\ 
        $\lambda$             & The amplification effect of local environment. 
                     \\ 
        $\delta$     & Cross-layer feedback strength from PGG to IG.                         \\ 
        $\eta$       & Cross-layer feedback strength from IG to PGG.                       \\ 
    \end{tabular}
    \end{ruledtabular}
\end{table}
\subsection{Payoff calculation in the PGG}
In the PGG layer, players choose either to contribute cost $c_2$ to a common pool (cooperate) or not (defect). Contributions are multiplied by a synergy factor $r$ and evenly distributed.

We incorporate feedback from the competitive IG layer through the effective synergy factor $r^{*}$. Intense competition on the IG layer can weaken the efficiency of cooperation on the PGG layer, and vice versa. We define the effective synergy factor for group $i$ at time $t$ as
\begin{equation}\label{feedback_2}
    r_{i}^{*}(t)=r\left[1-\eta\left(\frac{\lambda \mathcal{N}_{\mathrm{D}, i}^{[1]}(t)}{N}-1\right)\right].
\end{equation}
where $\eta$ is the cross-layer feedback strength from IG to PGG. At $\eta=0$, we have $r_{i}^{*}(t)=r$, and the PGG layer becomes the traditional public goods game model. At $\eta=1$, the effective synergy factor in the PGG completely depends on the feedback from the IG layer.

The payoffs for a player (C or D, layer 2) in group $i$ at time $t$ are calculated as
\begin{subequations}
    \begin{align}
        \pi_{\mathrm{C},i}^{[2]}(t)&=\frac{\mathcal{N}_{\mathrm{C},i}^{[2]}(t)c_2}N r_{i}^{*}(t)-c_2,\\
        \pi_{\mathrm{D},i}^{[2]}(t)&=\frac{\mathcal{N}_{\mathrm{C},i}^{[2]}(t)c_2}N r_{i}^{*}(t).
    \end{align}
\end{subequations}

Total payoffs of player $i$ in layer 2 are similarly calculated from multiple group interactions,
\begin{equation}\label{Pi_lay2}
    \Pi_i^{[2]}(t)=
        \begin{cases}
            \sum_{j\in\Omega_i^{[2]}}\pi_{\mathrm{C},j}^{[2]}(t),&\quad \mathrm{if}~ S_i^{[2]}(t)=1,\\[6pt]
            \sum_{j\in\Omega_i^{[2]}}\pi_{\mathrm{D},j}^{[2]}(t),&\quad \mathrm{if}~ S_i^{[2]}(t)=0.
        \end{cases}
\end{equation}

\begin{figure*}[htbp]
    \centering
    \includegraphics[width=1\textwidth]{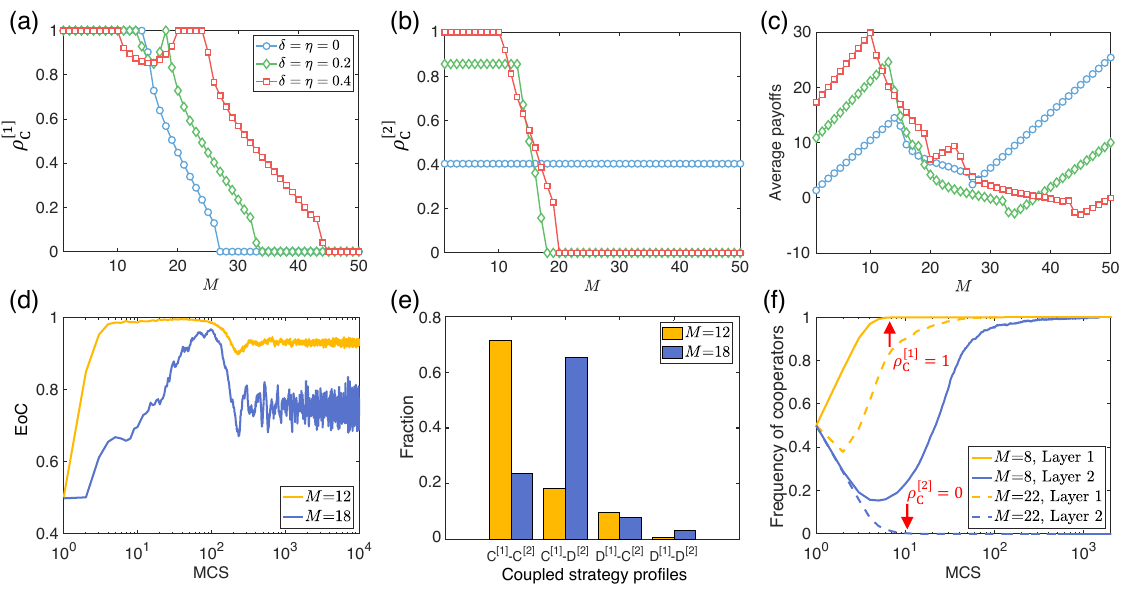}
	\caption{\textbf{Impact of baseline resources on the evolution of cooperation in two-layer networks}. Panels (a)--(c) show how cooperation fraction in the IG and PGG layers, along with the overall average payoffs, varies with $M$ under different cross-layer feedback strengths $\delta$ and $\eta$. Panels (d) and (e) depict the evolution of the Environment of Cooperators (EoC) and the fraction of coupled strategy profiles for $M = 12$ and $M = 18$ when $\delta = \eta = 0.4$. Panel (f) shows the time evolution of cooperation fraction in both layers for $M = 8$ and $M = 22$ under $\delta = \eta = 0.4$. The fixed parameters are $r = 3.7$ and $d = 6$.} 
	\label{hz_M}
\end{figure*}

\subsection{Strategy update rule}
All players update their strategies in each layer independently, following the pairwise comparison rule without mutation. For layer $l$, the probability that player $i$ accepts the strategy of the random selected neighbor $j$ is given by
\begin{equation}
    P_{S_i^{[l]}\leftarrow S_j^{[l]}}(t)=\frac1{1+\exp[(\Pi_i^{[l]}(t)-\Pi_j^{[l]}(t))/k]},
\end{equation}
where noise parameter $k>0$ controls update randomness~\cite{szabo1998evolutionary}. A smaller $k$ value corresponds to more greedy updates, whereas a larger $k$ value leads to more random updates.

The independence of strategy updates in the two-layer networks reflects the adaptability of players in cooperative and competitive environments. A player might change their PGG strategy while keeping their IG strategy, or vice versa. This allows individuals to reconsider their approach to cooperation without necessarily altering their competitive behavior.

\subsection{Simulation setup}
Unless stated otherwise, we initialize simulations with each player randomly choosing C or D on each layer with equal probability. We have verified that varying the initial fraction of cooperators does not qualitatively affect the eventual outcome. We use population size $L^2 = 200 \times 200$ for sufficient sample size. We employ a synchronous Monte Carlo simulation in which each player on each layer updates its strategy exactly once per basic Monte Carlo step (MCS). Each simulation is run for a sufficiently long time (typically $10^4$--$10^6$ MCSs) to ensure the population reaches a steady state. We then measure averages over a further $10^3$ MCSs to obtain stable statistics. To reduce noise, results for each set of parameters are averaged over at least 10 independent simulation runs with different random initial conditions.

The main symbols we use in this work are listed in Table~\ref{tab:symbols}. To focus on the effects of cross-layer feedback, we fix the parameters $c_1 = c_2 = 1$ and $k = 0.1$ for convenience. We also set $\beta = 1$ unless exploring its effect, meaning that any advantage to defectors in IG comes solely from their effort. Additionally, we assume $\lambda = 2$, so that if the fraction of cooperators in a group exceeds $1/2$, the other layer's group receives a positive feedback, and vice versa. We verify the robustness of $d$, $\beta$, and $\lambda$ in Fig.~\ref{figa1} and find that changing the values of these three parameters does not affect our conclusions qualitatively.

\begin{figure*}[htbp]
    \centering
    \includegraphics[width=1\textwidth]{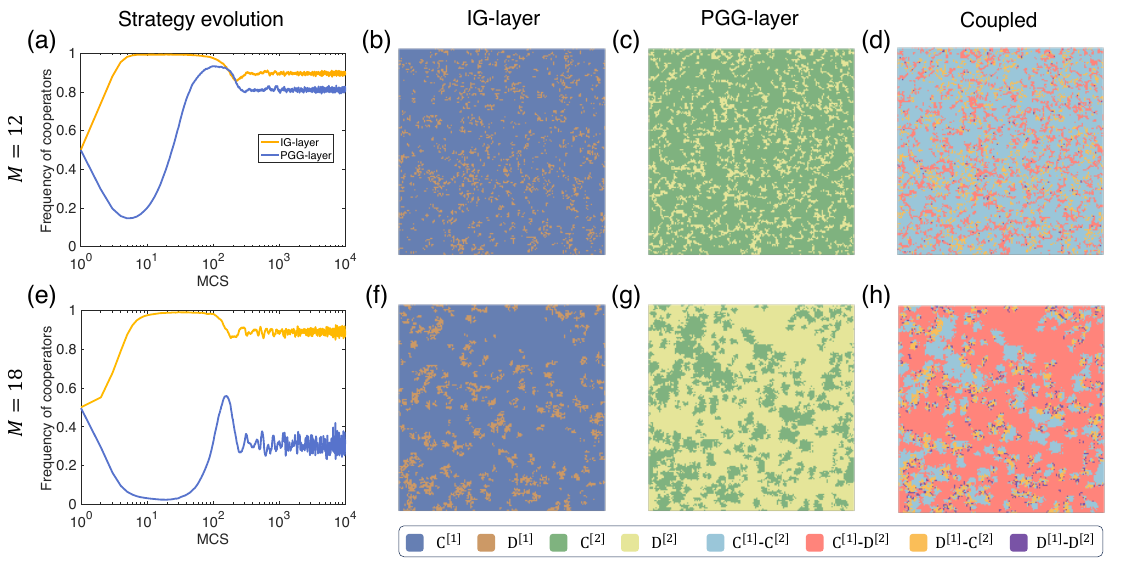}
	\caption{\textbf{Aggregation of IG defectors suppresses PGG cooperation}. We compare the spatio-temporal evolution of strategies in the two-layer network. The first and second rows correspond to $M = 12$ and $M = 18$, respectively. The first column presents cooperation frequency in both layers as a function of MCS. The second and third columns show snapshots of the strategy distributions in the IG and PGG layers, along with the coupled strategies at the evolutionary equilibrium. Different colors distinguish cooperative and defective strategies in each layer and their coupled forms. Fixed parameters: $r = 3.7$, $d = 6$, $\delta = \eta = 0.4$.} 
	\label{kz}
\end{figure*}

\begin{figure*}[htbp]
    \centering
    \includegraphics[width=1\textwidth]{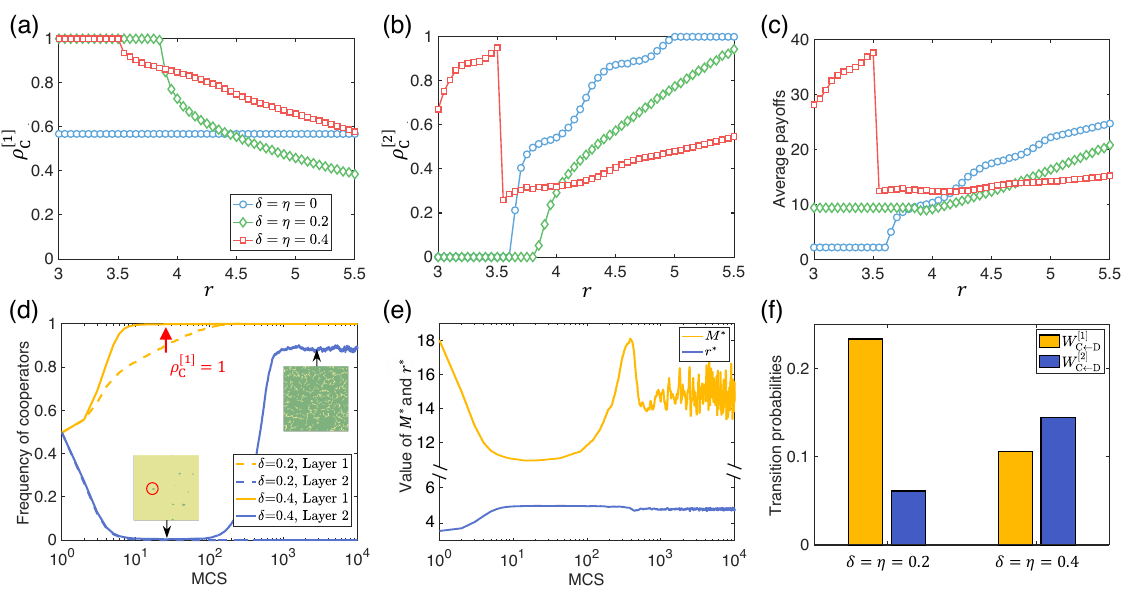}
	\caption{\textbf{Influence of synergy factor on cooperation evolution}. Panels (a)--(c) show how cooperation fraction in the IG and PGG layers, along with the overall average payoffs, changes with $r$ under different cross-layer feedback strengths $\delta$ and $\eta$. Panel (d) illustrates the evolution of cooperation fraction in both layers for $r = 3.3$ under two distinct feedback strengths. The insets in panel (d) give snapshots of the PGG layer’s strategy distributions during its initial decline phase and its evolutionary stable phase for $\delta = \eta = 0.4$, with green representing cooperators and yellow representing defectors. Panel (e) displays the time evolution of the average $M^*$ and $r^*$ for $r = 3.55$ under $\delta = \eta = 0.4$. Panel (f) presents the average probability of cooperators switching to defectors in both layers for $r = 5$ under $\delta = \eta = 0.2$ and $0.4$. The fixed parameters are $M = 18$ and $d = 6$.}
	\label{hz_r}
\end{figure*}

\begin{figure*}[htbp]
    \centering
    \includegraphics[width=1\textwidth]{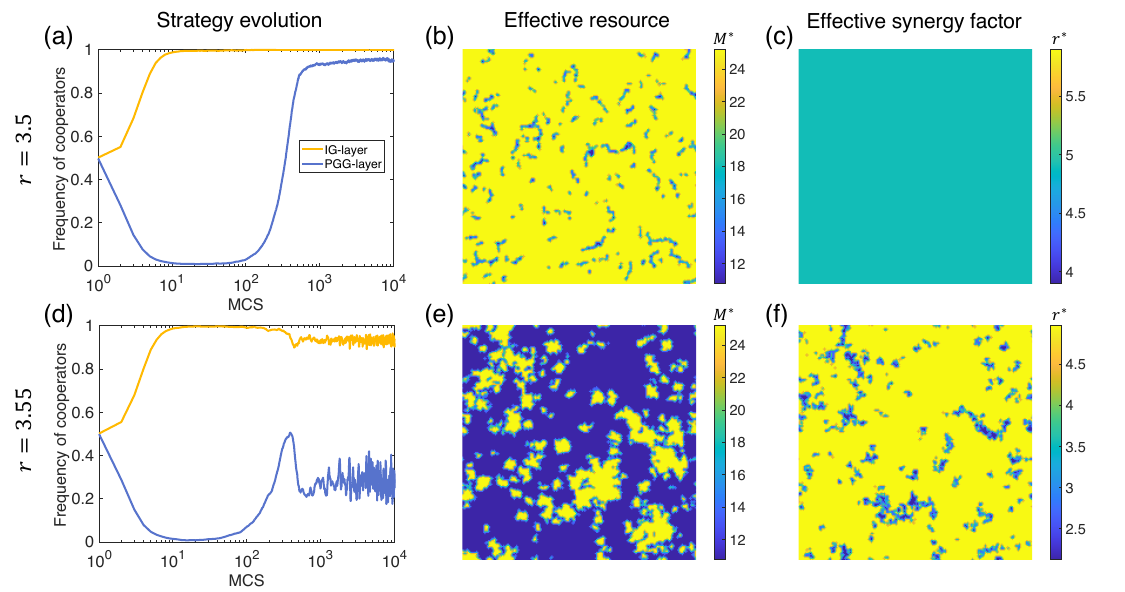}
	\caption{\textbf{Discontinuous PGG transitions due to increased synergy factor}. The first and second rows depict the evolutionary outcomes for $r = 3.5$ and $r = 3.55$, respectively. In each row, the left column shows the time evolution of cooperation frequency in both layers, while the middle and right columns present snapshots of the spatial distributions of effective resource $M^*$ and synergy factor $r^*$ at evolutionary equilibrium. The fixed parameters are $M = 18$, $d = 6$, and $\delta = \eta = 0.4$.} 
	\label{tran}
\end{figure*}

\section{Results}\label{sec3}

\subsection{Emergence of six distinct states as $M$ increases}

We first investigate how changes in the baseline resource $M$ in the IG layer influence cooperative dynamics across both layers. Intuitively, $M$ controls the level of rewards in the IG. If $M$ is very small, there is little to fight over, which might discourage costly competition; if $M$ is large, the stakes of cooperation are high, which may encourage defections in layer 1. Our simulations confirm that increasing $M$ drives the system through a series of phase transitions between distinct cooperation/defection states on the two layers.

As shown in Figs~\ref{hz_M}(a)--(c), the two-layer system transitions among six qualitatively distinct states as $M$ increases under cross-layer feedback. For small $M$, the population settles in either $(\mathrm{C},\mathrm{C})$ or $(\mathrm{C}, \mathrm{C+D})$ states, where the notation $(\mathrm{X},\mathrm{Y})$ describes pipulation states with $\mathrm{X}$ in the IG layer and $\mathrm{Y}$ in the PGG layer. In both cases, players have little incentive to expend the higher effort because the resource $M$ is too meager to justify the cost $d$. This peaceful competition on layer 1 in turn provides positive feedback to layer 2, and the abundance of cooperators on layer 2 further increases the available resources on layer 1 via $\delta$-feedback, creating a virtuous cycle that boosts the overall payoffs to the population(Fig.~\ref{hz_M}(c)).

As $M$ increases to moderate values, the system shifts into a mixed-mixed state $(\mathrm{C+D}, \mathrm{C+D})$ and the average payoffs reache a local optimum. The IG layer now has some incentive for defection and the presence of these IG defectors starts to suppress the synergy in the PGG layer, so the PGG layer can no longer maintain full cooperation. Notably, similar competition levels in the IG layer can lead to different outcomes in the PGG layer due to varying local environments. To quantify this effect, we introduce the Environment of Cooperators (EoC):
\begin{equation}
    \text{EoC}(t) = \frac{\sum_{i=1}^{L^2} S_i^{[1]}(t) \cdot \mathbf{1} ( \sum_{j \in \Omega_i} S_j^{[2]}(t) > \frac{\beta}{N})}{\sum_{i=1}^{L^2} S_i^{[1]}(t)}.
\end{equation}
where $\mathbf{1}(\cdot)$ is an indicator function. Higher EoC values reflect a more cooperative-friendly environment in the PGG layer. As illustrated in Fig.~\ref{hz_M}(d), for $M = 18$, cooperators in the PGG layer encounter a harsher competitive environment, evidenced by an increased concentration of defectors in the IG layer (Figs.~\ref{kz}(b) and (f)). That is to say, the aggregation of defectors in the IG layer will undermine the cooperation in the PGG layer (Figs.~\ref{kz}(c) and (g)). Consequently, in Fig.~\ref{hz_M}(e) and Figs.~\ref{kz}(d) and (h), the type of coupled strategy profiles changes from $\mathrm{C^{[1]}}$-$\mathrm{C^{[2]}}$ to $\mathrm{C^{[1]}}$-$\mathrm{D^{[2]}}$.

Once cooperation collapses in the PGG layer, the IG layer may paradoxically revert to full cooperation. Despite positive feedback from IG to PGG, the mismatch in feedback timing can lock the PGG layer into persistent defection (Fig.~\ref{hz_M}(f)). As $M$ increases further, the population passes through $(\mathrm{C+D}, \mathrm{D})$ and eventually reaches full defection $(\mathrm{D}, \mathrm{D})$. Under no feedback, cooperators in the IG layer would vanish at $M' \approx 26.6$, whereas under feedback they disappear at $M'/(1-\delta)$, creating a local payoff minimum. Although average payoffs rise slowly as $r$ increases after $(\mathrm{D}, \mathrm{D})$ takes over, the population loses the synergy benefits of the PGG layer as well as reduces the available resources in the IG layer, leading to a lower social welfare compared to the no-feedback scenario.

\subsection{Increasing $r$ promotes cooperation but escalates competition}
Next, we explore how variations in the baseline synergy factor $r$ in the PGG layer influence cooperative dynamics. Figures~\ref{hz_r}(a)--(c) show that smaller $r$ values maintain full cooperation in the IG layer, whereas cooperation in the PGG layer either collapses to full defection under weak feedback or stabilizes at intermediate levels under stronger feedback. Temporal dynamics (Fig.~\ref{hz_r}(d)) show weak feedback strength ($\delta=0.2$) at $r=3.3$ leading to premature cooperation collapse in PGG, whereas strong feedback strength ($\delta=0.4$) preserves cooperation due to an amplified synergy factor, enabling reciprocity-driven stabilization of cooperative clusters~\cite{perc2008restricted,szolnoki2009promoting}.

As $r$ reaches a critical threshold under stronger feedback ($\delta = \eta = 0.4$), cooperation in the PGG layer undergoes a discontinuous phase transition (Fig.~\ref{hz_r}(b)). Figure~\ref{tran} explains this phenomenon in terms of the evolution of the strategies and the spatial distribution of the available resources and the actual synergy factors. At $r=3.5$, the PGG layer remains in a low-cooperation state until the IG layer achieves full cooperation, creating a favorable environment that deterministically promotes PGG cooperation. In contrast, at $r=3.55$, PGG cooperators start increasing after an initial decline, driven by partial cooperation growth in the IG layer which temporarily elevates the synergy factor $r^*$ (Fig.~\ref{hz_r}(e)). However, this cooperation resurgence in the PGG layer boosts resource availability in the IG layer, subsequently fostering IG defectors and causing renewed suppression of PGG cooperation. Ultimately, $r^*$ stabilizes around 4.52 (Fig.~\ref{hz_r}(e)). Although this average $r^*$ is much higher than the baseline $r$, regions with low $r^*$ values hinder further expansion of cooperator clusters (Fig.~\ref{tran}(f)).

Further increases in $r$ continue to favor PGG layer's cooperation while increasing IG layer’s competition. Moreover, for larger $r$, raising the feedback strength increases the fraction of cooperation in the IG layer but suppresses cooperation in the PGG layer. To evaluate these steady-state dynamics, we measure the probability that cooperators switch to defectors as
\begin{equation}
\resizebox{.9\columnwidth}{!}{$
W_{\mathrm{C} \leftarrow \mathrm{D}}^{[l]} = \sum_{t=T-\tau+1}^{T} \frac{\sum_{i=1}^{L^2} S_{i}^{[l]}(t) \left(1 - S_{j}^{[l]}(t)\right) P_{S_{i}^{[l]} \leftarrow S_{j}^{[l]}}(t)}{\tau \sum_{i=1}^{L^2} S_{i}^{[l]}(t)}.$}
\end{equation}
where $j \sim \mathcal{U}(\Omega_i^{[l]}\setminus \{i\})$ is a randomly chosen neighbor of $i$, $T$ is the total number of MCS, and $\tau$ is the averaging window. As shown in Fig.~\ref{hz_r}(f), stronger feedback lowers the chance of IG-layer cooperators switching to defection but increases this probability in the PGG layer. Consequently, the IG layer retains more cooperators, while the PGG layer experiences a decline in cooperation.
\begin{figure*}[htbp]
    \centering
    \includegraphics[width=1\textwidth]{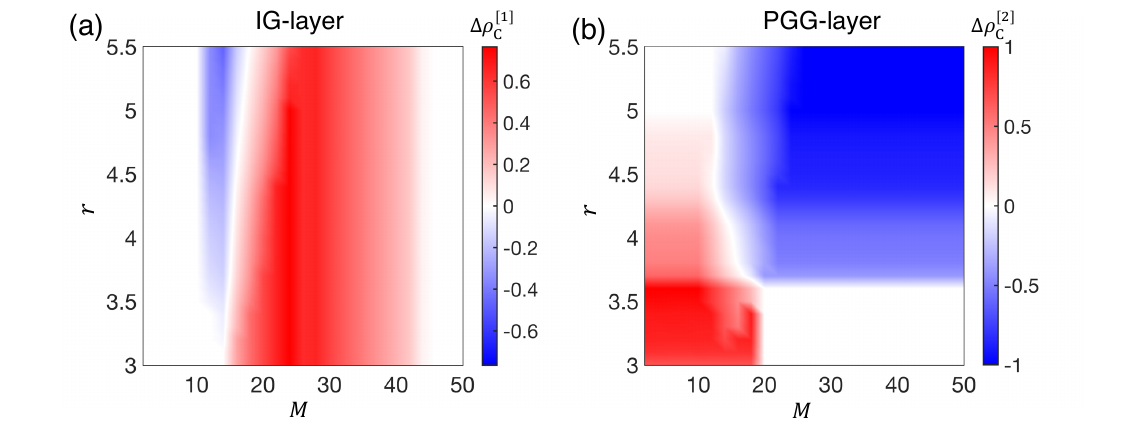}
    \caption{\textbf{Impact of coupled feedback on cooperation in $M$-$r$ parameter space.} Panels (a) and (b) show heatmaps of the differences in cooperation fraction in the IG and PGG layers, respectively, comparing the coupled-feedback scenario to the no-feedback baseline. Red denotes regions where feedback promotes cooperation, blue regions denote suppression of cooperation, and white regions indicate no change. Fixed parameters are $d = 6$, $\delta = \eta = 0.4$.}
    \label{hm_Mr}
\end{figure*}

\begin{figure*}[htbp]
    \centering
    \includegraphics[width=1\textwidth]{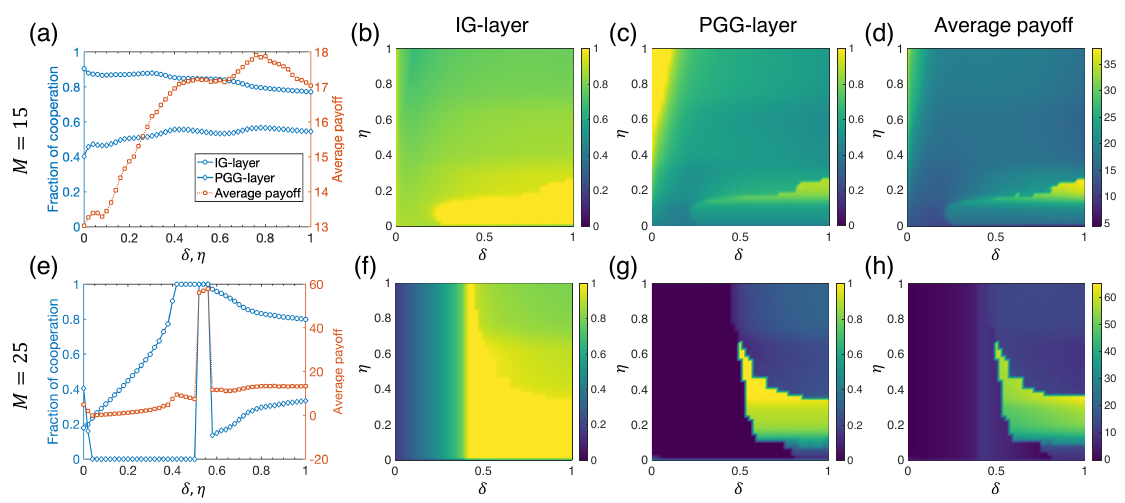}
    \caption{\textbf{Optimal feedback strengths maximizing cooperation.} The first and second rows correspond to $M = 15$ and $M = 25$, respectively. The first column illustrates the effects of homogeneous feedback strength on the fraction of cooperation and the overall average payoff within the two-layer network. Columns two through four display heatmaps that depict the influence of heterogeneous feedback strengths on the IG layer (panels (b) and (f)), the PGG layer (panels (c) and (g)), and the overall average payoff (panels (d) and (e)), respectively. The fixed parameters are $r=3.7$ and $d = 6$.}
    \label{delta_eta}
\end{figure*}
\subsection{Cross-layer feedback has complex impacts in the $M$-$r$ parameter space}
Having examined the effects of $M$ and $r$ on the evolution of cooperation separately, we now turn to the combined effect of the cross-layer feedback mechanism across the $M$-$r$ parameter space. We are concerned about under what conditions does coupled feedback between layers improve the prevalence of cooperation, and when does it hurt? To answer this, we compare outcomes in the coupled feedback scenario (finite $\delta, \eta$) with a baseline scenario without feedback ($\delta=\eta=0$). Rather than uniformly facilitating cooperation, we find that the feedback can both enhance or inhibit cooperation depending on the parameters.

Figure~\ref{hm_Mr} shows heatmaps of difference $\Delta \rho_{\mathrm{C}}^{[l]}$ between the fraction of cooperarion in each layer with and without feedback. When $M$ is extremely small or extremely large, the feedback has negligible impact on the IG layer. If $M$ is very small, IG layer is going to be cooperative with or without feedback, so adding feedback doesn’t change that much. If $M$ is very large, then the IG layer is destined to be highly competitive regardless of feedback, so again feedback doesn’t change the fate of the IG. In an intermediate $M$ range, however, the system shifts from suppressing to promoting IG cooperation as $M$ grows, and larger $r$ pushes this crossover to higher $M$. 

In the PGG layer, at smaller $M$ (particularly when $M<20$), low-to-moderate $r$ values amplify the cooperative effects of feedback. However, as $r$ grows in that small-$M$ scenario, the promotion effects saturates or can even invert. Meanwhile, for large $M$, the PGG layer tends to lose cooperators due to feedback at high $r$ (blue region in upper-right of Fig.~\ref{hm_Mr}(b)). This corresponds to the case we see where abundant resources and high synergy produce overly intense IG competition, which then weakens PGG cooperation (the negative $\eta$ feedback dominates).

In summary, the effect of cross-layer feedback is double-edged. In certain regimes (e.g., moderate $M$ but not super high $r$), the feedback provides a needed boost. But in other regimes (e.g., large $r$ or large $M$), the feedback creates an imbalance, each layer’s potential cooperative gains are offset by heightened defection in the other layer, yielding no net improvement over an uncoupled scenario.

\subsection{Optimizing cross-layer feedback for maximal cooperation}
Finally, we investigate how varying the feedback strengths $\delta$ and $\eta$ affects evolutionary results, particularly whether there is an optimal way to tune the coupling. In all previous analyses, we have assumed symmetric or fixed feedback. Now we ask: if we can choose the degree of coupling in each direction, what combination best promotes cooperation and overall social welfare?

Under homogeneous feedback conditions at smaller $M$ (Fig.~\ref{delta_eta}(a)), simultaneously increasing the homogeneous feedback slightly reduces cooperation in the IG layer but enhances it in the PGG layer. Moreover, there exists an optimal homogeneous feedback strength that maximizes the overall average payoff. At larger $M$ (Fig.~\ref{delta_eta}(e)), the fraction of cooperation in the IG layer initially rises and then declines as $\delta$ and $\eta$ increase, with a moderate feedback range yielding full cooperation. However, the PGG layer loses cooperators at a weak feedback strength due to the intensely competitive IG environment. Once the feedback strength reaches a certain moderate level, the PGG layer undergoes a discontinuous transition from full defection to full cooperation, driven by the IG layer’s rapid convergence to full cooperation status and the fully amplified synergy factor, ultimately maximizing the population’s overall payoff. Hence, for larger $M$, there exists an optimal homogeneous feedback strength that simultaneously maximizes cooperation and payoff. Beyond this point, defectors emerge in the IG layer, and the PGG layer shifts from full cooperation to a low-cooperator coexistence. Further increases in the homogeneous feedback strength continue to decrease the fraction of cooperation in the IG layer and increase it in the PGG layer.

Under heterogeneous feedback, we find that an imbalance in feedback strengths can sometimes be beneficial. For the case of smaller $M$ (Figs.~\ref{delta_eta}(b)--(d)), two distinct optimal parameter regions emerge: one characterized by small $\delta$ and large $\eta$, and another with large $\delta$ and small $\eta$, each facilitating cooperation and high overall payoffs. This suggests that pronounced feedback disparities between layers can simultaneously enhance cooperation. For larger $M$ values (Figs.~\ref{delta_eta}(f)--(h)), the benefits of heterogeneity become more conditional. Small $\delta$ yields low payoffs regardless of $\eta$, due to persistent defection in the PGG. Conversely, large $\delta$ values coupled with moderate $\eta$ sustain full cooperation across both layers and maximize overall payoffs. These insights highlight the importance of tuning feedback strengths to achieve optimal cooperative outcomes across diverse conditions.

\section{Discussion}\label{sec4}
Understanding the interplay between cooperation and competition is essential for explaining collective behaviors and social welfare in complex systems. Traditional evolutionary game models have studied cooperation and competition in isolation or in simplified combinations~\cite{wang2021replicator,wang2022involution,wang2022modeling,huang2024memory,lyu2024evolution}. However, real-world systems contain complex feedbacks between cooperation and competition. For instance, cooperation in one domain can increase effective resource in another and promote competition, while intensive competition in turn suppresses cooperation. This work aims to fill this gap by developing a multilayer network model~\cite{wang2015evolutionary} that captures both cooperative and competitive interactions, investigating how cross-layer feedback influences the evolution of cooperation and competition.

We model the interplay between cooperation and competition in a two-layer network, where one layer corresponds to a competitive environment and another serves as a cooperative environment. We use the involution game~\cite{wang2021replicator} to model competitive interactions and the public goods game~\cite{santos2008social} to denote cooperative interactions. In the presence of cross-layer feedback, as the baseline resource $M$ increases, we find a rich array of collective behaviours, ranging from full cooperation to full defection with a variety of mixed states in between. These phase transitions reveal how resource availability can impact cooperative dynamics in coupled competitive-cooperative systems. It also indicates that in multidomain systems, simply increasing resources can backfire, promoting excessive competition and ultimately weakening cooperation.

Our analyses show that moderate $M$ and $r$ can promote social welfare, especially when feedback strength is large. However, too large $M$ or $r$ leads to intensified competition and erodes the conditions for cooperation. Notably, we observe critical $r$ thresholds for discontinuous phase transitions in cooperation fraction, especially in the PGG layer with strong feedback. By exploring the effect of cross-layer feedback in the $M$-$r$ parameter plane, we find that its effect on the coupled systems is double-edged. On the one hand, net positive effects occur where one layer struggles to sustain cooperation but the other provides support. On the other hand, net negative effects occur where one layer’s strong cooperative drive sparks an overly intense competitive response in the other. Only a narrow band of moderate $M$ and smaller $r$ can sustain cooperation in both layers. These findings highlight that more coupling is not always better and the baseline conditions matter greatly.

Moreover, we find that the existence of optimal cross-layer feedback promotes overall cooperation and maximizes social welfare. While homogeneous feedback strength can support cooperation under certain conditions, our results show that tuning distinct feedback intensities for resource availability and synergy effectiveness could better facilitate overall cooperation and average payoffs. Our exploration of the $\delta$-$\eta$ space reveals that sometimes an imbalanced coupling (one direction is strong, the other is weak) can promote cooperation on both layers, especially when baseline resources are not large. This provides a perspective for designing interventions in real systems where greater feedback between competition and cooperation is not always better, and where targeted integration, perhaps even one-directional, may produce the best results.

In the current context of economic growth, intensifying competition manifests across multiple scales: individual-level involution, corporate rivalry, and inter-state confrontations exemplified by tariff wars or even local wars. The need to establish cooperative and win-win strategies has never been more compelling or urgent. In this study, we formalize feedback mechanisms between competition and cooperation. Crucially, we demonstrate feasible pathways by which cooperative interactions can amplify the collective social welfare and thereby mitigate the harms wrought by excessive competition. Our framework also reveals great potential for transforming zero-sum competition into positive-sum coevolution, providing valuable insights into promoting collective cooperation in an otherwise highly competitive world.

Future studies can extend our model by considering more realistic features such as dynamic network structures~\cite{su2023strategy}, mutation-driven strategy diversification~\cite{allen2012mutation}, and stochastic~\cite{zhu2024evolutionary}. In addition, our framework allows for the introduction of external factors such as collective risk~\cite{hua2024coevolutionary}, reciprocity mechanisms~\cite{ma2024effect}, etc. to explore their effects on cross-layer feedback. By developing specific strategies, cooperation can be promoted in the presence of unavoidable competitive forces.

\begin{acknowledgments}
This work is supported by National Science and Technology Major Project (2022ZD0116800), Program of National Natural Science Foundation of China (12425114, 62141605, 12201026, 12301305), the Fundamental Research Funds for the Central Universities, and Beijing Natural Science Foundation (Z230001).
\end{acknowledgments}

\appendix
\setcounter{figure}{0}
\renewcommand{\thefigure}{A\arabic{figure}}

\begin{figure*}[htbp]
    \centering
    \includegraphics[width=1\textwidth]{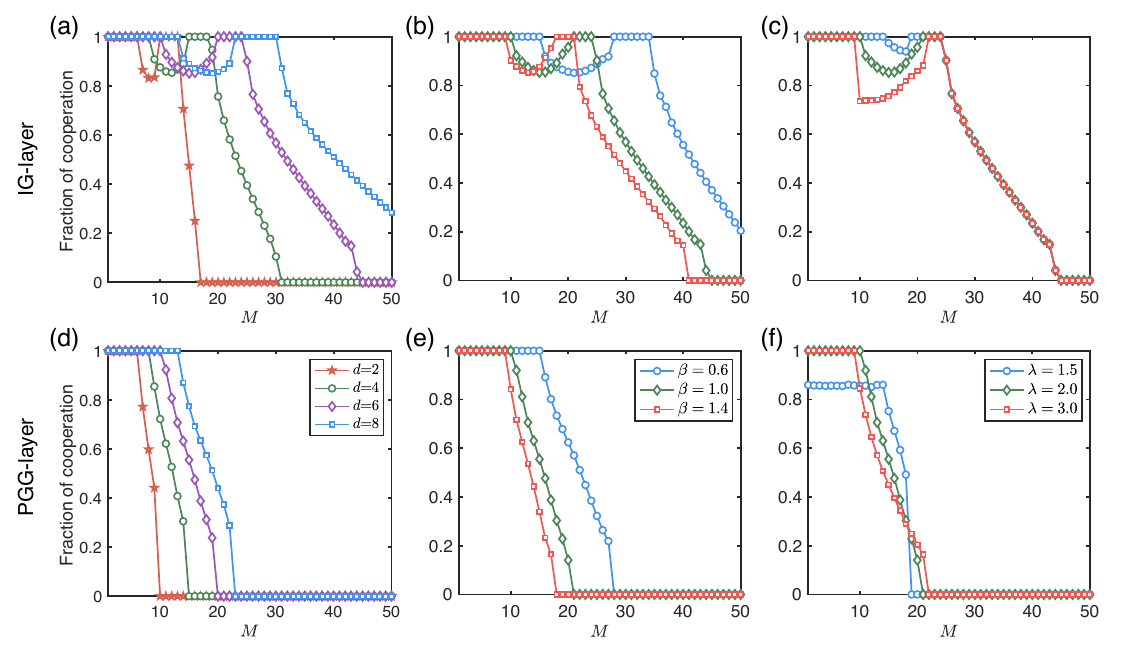}
	\caption{\textbf{Robustness verification with different parameters}. The first and second rows show the evolution of cooperation in the IG and PGG layers, respectively. Columns—left to right: panels (a) and (d) the defector cost in the IG layer, $d$; panels (b) and (e) the defectors’ relative advantage in the IG layer, $\beta$; panels (c) and (f) the amplification effect of cooperation, $\lambda$. All other parameters are fixed at $r = 3.7$ and $\delta = \eta = 0.4$.} 
	\label{figa1}
\end{figure*}

\section*{Appendix: Robustness verification}\label{appendix}
In this section, we verify that for different parameters $d$, $\beta$ and $\lambda$, there is no change in the qualitative conclusions.

We first study how variations in the defector cost $d$ modulate the coupled dynamics. Figures~\ref{figa1} (a) and (d) reveal a non-monotonic response on the IG layer, where the fraction of cooperation initially declines as the $M$ increases, then recovers, and finally decays again. On the PGG layer the pattern is simpler, and all trajectories decrease monotonically from full cooperation to full defection. While changing $d$ leaves these qualitative trends unchanged, it changes the quantitative thresholds. A larger $d$ triggers the first appearance of defectors at a smaller $M$ and drives both layers to complete defection at lower $M$ values. The same logic emerges for the defectors’ relative advantage $\beta$ (Figs.~\ref{figa1} (b) and (e)). Reducing $\beta$ systematically delays the breakdown of cooperation, confirming that a weaker competitive edge for defectors facilitates cooperative persistence in the cross-layer system.

Figures~\ref{figa1} (c) and (f) examine the amplification effect of cooperation $\lambda$. A larger $\lambda$ lowers the $M$ threshold at which cooperation begins to decrease, indicating that the lower the configuration of local cooperators required for positive feedback, the easier it is to weaken cooperation with smaller base resources. Once $M$ is sufficiently large, at this point the PGG layer has collapsed to full defection state and changes in $\lambda$ no longer affect the IG layer. At the PGG layer itself, a threshold value of $M$ distinguishes between two situations, below which a smaller $\lambda$ favours cooperation and above which a larger $\lambda$ becomes favourable.

\bibliography{ref}

\end{document}